# 基于光场幅度动态变化的 OCT 功能成像原理和应用

杨建龙 [1,*]，张浩然 [1]，刘畅 [1]，谷成富 [1]

[1] 上海交通大学生物医学工程学院，上海 200240

**摘 要** 光学相干层析成像（OCT）能够无损获得微米级空间分辨率的样品截面信息，在眼科、血管内科等临床诊疗研究和应用起到了重要的作用。通过 OCT 测量光场的幅度可以获得样本的三维结构信息，如眼底视网膜的分层结构，但对于组织特异性、血流、力学特性等功能信息的作用有限。基于相位、偏振态、波长等光场参量的 OCT 功能成像技术应运而生，如多普勒 OCT、OCT 弹性成像、偏振敏感 OCT、可见光 OCT 等。其中，基于光场幅度动态变化的 OCT 功能成像技术具有显著的鲁棒性和系统复杂度优势，已经在临床无标记血管造影中获得成功。此外，应用于三维血流流速测量的动态光散射 OCT，具有无标记组织/细胞特异性显示能力的动态 OCT，能够监控热物理治疗温度场的 OCT 温度层析成像等，已经成为了 OCT 功能成像的技术前沿。本文综述了基于光场幅度动态变化的 OCT 功能成像技术的原理、实现和应用，分析了所面临的技术挑战，并展望了未来发展方向。

# Dynamic Change of Amplitude for OCT Functional Imaging

Yang Jianlong[1,*], Zhang Haoran[1], Liu Chang[1], Gu Chengfu[1]

[1] *School of Biomedical Engineering, Shanghai Jiao Tong University, Shanghai 200240, Shanghai*

**Abstract** Optical coherence tomography (OCT) is capable of non-destructively obtaining cross-sectional information of samples with micrometer spatial resolution, which plays an important role in ophthalmology and endovascular medicine. Measuring OCT amplitude can obtain three-dimensional structural information of the sample, such as the layered structure of the retina, but is of limited use for functional information such as tissue specificity, blood flow, and mechanical properties. OCT functional imaging techniques based on other optical field properties including phase, polarization state, and wavelength have emerged, such as Doppler OCT, optical coherence elastography, polarization-sensitive OCT, and visible-light OCT. Among them, functional imaging techniques based on dynamic changes of amplitude have significant robustness and complexity advantages, and achieved significant clinical success in label-free blood flow imaging. In addition, dynamic light scattering OCT for 3D blood flow velocity measurement, dynamic OCT with the ability to display label-free tissue/cell specificity, and OCT thermometry for monitoring the temperature field of thermophysical treatments are the frontiers in OCT functional imaging. In this paper, the principles and applications of the above technologies are summarized, the remaining technical challenges are analyzed, and the future development is envisioned.

**Key words** optical coherence tomography; functional imaging; dynamic change of amplitude; dynamic light scattering

**基金项目：**国家自然科学基金（62105198）

**通信作者：**[*]E-mail：jyangoptics@gmail.com

# 1 引　言

光学相干层析成像（Optical Coherence Tomography, OCT）利用时间相干性对背向散射光子进行筛选，能够非侵入、无损地获得被探测样品（如生物组织、工业材料等）的截面信息，同时具有高分辨率（可达 1 微米）、高灵敏度（90 到 110 dB,取决于测量方法）等优势[1,2]。经过了三十多年的持续发展，OCT 已经成为与超声、核磁共振、CT 等并列的医学层析（tomographic）影像方法，并由于其分辨率优势，在眼科、血管内科等临床诊疗中发挥着独特作用[3–6]。

通过背向散射光场的幅度可以获得被测样品的三维结构信息，如眼底视网膜的分层结构[7]，但对于组织特异性、血流、力学特性等功能信息的作用有限。为实现基于 OCT 的功能成像，研究者们将目光投向的 OCT 探测光场的其它维度（相位、波长、偏振态等）。Zhongping Chen 等人发展了多普勒（Doppler） OCT,利用生物组织中血流和 OCT 探测信号间的多普勒效应导致的相位变化，实现了精确的血液动力学测量[8,9]。同样基于相位变化，但是由于外界对组织的机械刺激，David D Sampson 等人发展出了高灵敏度 OCT 弹性成像，用于表征组织的生物力学性质[10,11]。Johannes F. de Boer 等人通过同时测量散射信号的两个正交偏振态（偏振敏感 OCT），实现了对生物组织中纤维状结构的特异性显示[12,13]，并成功用于热治疗过程的表征。Hao F. Zhang 等人发展了可见光 OCT，利用血液在 550 纳米波段的吸收特性，实现了对血氧饱和度的精确测量[14,15]。

然而，上述技术在实现不同 OCT 功能成像的同时也增加了系统的复杂度和成本。多普勒 OCT 和高灵敏度 OCT 弹性成像需要相位稳定的 OCT 系统[16]，还需要处理相位包裹（wrapping）等问题[17,18]。偏振敏感 OCT 需要额外的偏振分光和探测器件，并需要处理偏振模式色散等问题[13]。可见光 OCT 需要用在可见光波段低时间相干性、高空间相干性光源，目前的主要选项是超连续谱光源，其具有较高的成本和较强的噪声[19,20]。

另一方面，在 OCT 探测光场幅度（强度）用于生物组织结构成像基础之上，其动态变化，即时间维度信息，也被用于探索发展 OCT 功能成像技术。相较于相位、偏振态等测量量，基于光场幅度动态变化的功能成像技术在鲁棒性和系统复杂度方面具有显著优势。其中最为成功的案例是 OCT 血流成像（OCT angiography, OCTA）,分光谱幅度去相关（SSADA）[21]、散斑方差（svOCT）[22]、相关映射（cmOCT）[23]等算法已经被部署于商业化仪器，在临床研究和诊疗中得到了广泛的应用[24,25]。

此外，一些其它基于光场幅度动态变化的 OCT 功能成像方法也正在蓬勃发展，是 OCT

功能成像的技术前沿。其中主要包括：应用于三维速度测量的动态光散射 OCT（dynamic light scattering OCT, DLS OCT）[26]，具有无标记组织/细胞特异性显示能力的动态 OCT（dynamic OCT）[27],能够监控热物理治疗温度场的 OCT 温度层析成像[28]等。图 1 给出了基于光场幅度动态变化的 OCT 功能成像的应用示例，包括（a）血流成像[29]；（b）三维速度测量[30]；（c）组织特异性成像[31]；（d）温度层析成像[28]。

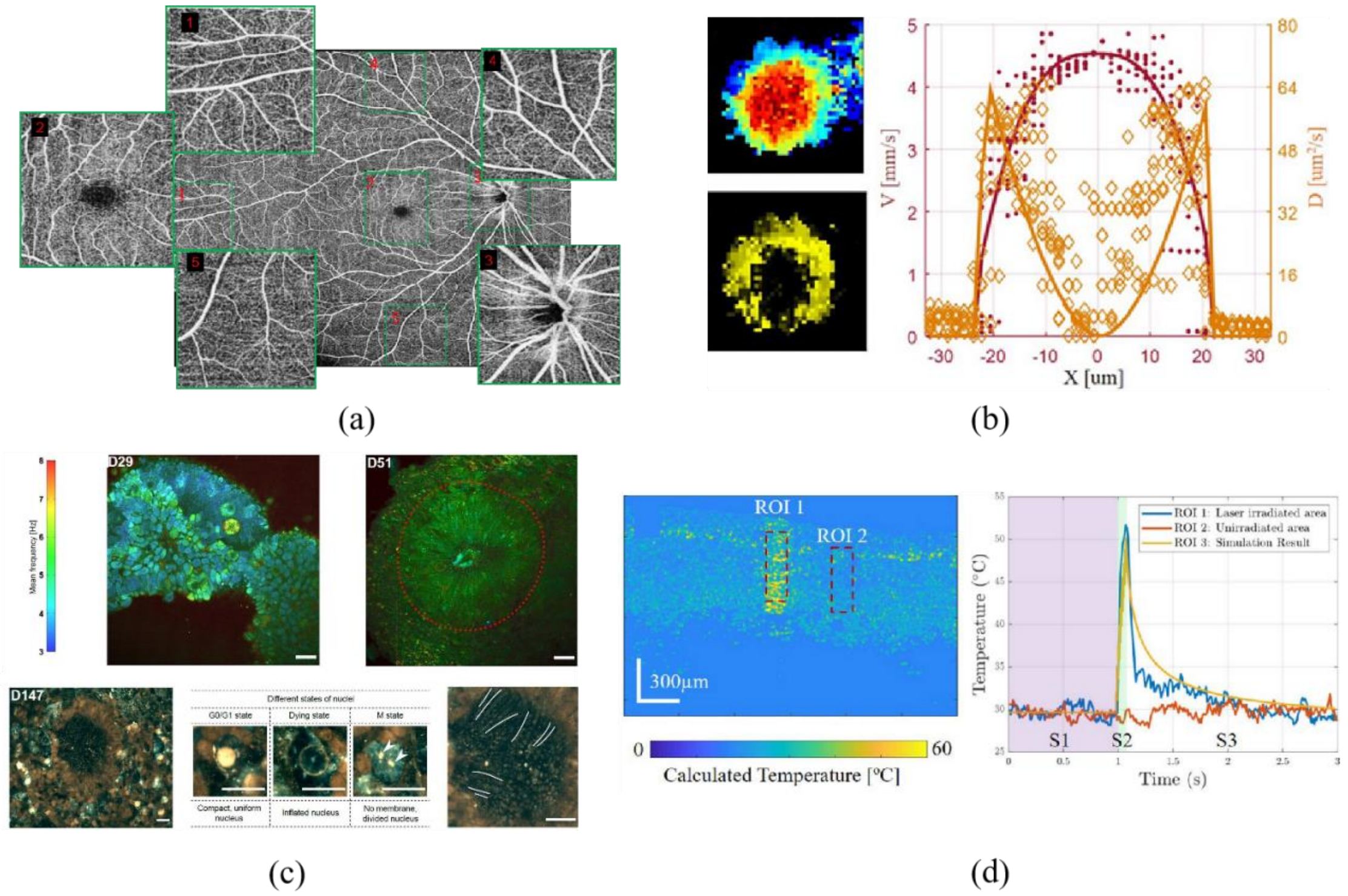

图1 光场幅度动态变化的OCT功能成像的应用示例。（a） 血流成像[29]；（b） 三维速度测量[30]；（c） 组织特异性成像[31]；（d） 温度层析成像[28]。

Fig. 1 Applications of dynamic change of amplitude in OCT functional imaging. (a) angiography[29]; (b) 3D velocimetry[30]; (c) tissue-specific imaging[31]; and (d) thermometry[28].

本文对基于光场幅度动态变化的 OCT 功能成像原理和应用进行了综述，首先从原理出发对上述各类方法进行了总结和归纳，然后介绍了每种功能成像方法的具体技术实现和应用，最后分析了基于光场幅度动态变化的 OCT 功能成像所面临的技术挑战并展望了其未来发展方向。相较于现有相关综述从特定功能成像技术或应用场景出发[24,25,32–35]，本文从上述各种技术背后共性的物理机制和方法出发，进行了综合性地总结和分析，对 OCT 和更大范围的光电信息领域的研究人员具有一定参考价值。

## 2 光场幅度动态变化用于 OCT 功能成像的物理机制

生物体内的微观运动涵盖了多个层面的动态活动[36]。其中包括血液在血管中的流动，通

过心脏的跳动和血管的舒张与收缩，将氧气、养分输送到全身各部位，同时将二氧化碳等废物运送到一定的部位排出体外。此外，细胞内也存在着许多微观运动，如细胞器在细胞质中的移动，以维持能量供应和代谢平衡，以及细胞骨架的重组，使细胞得以适应不同的环境和任务[37]。分子水平上，生物分子持续地进行着交互和运动，参与各种生化反应和信号传递。蛋白质在生物体内也进行着构象变化和相互作用，实现多种功能，包括催化化学反应和细胞通信[38]。细胞之间的相互作用包括细胞的迁移、聚集和信号传导，这对于组织的发展、免疫系统的功能以及伤口修复都至关重要。此外，细胞周期中的运动确保了遗传信息的传递和维护[39]。

从物理模型的角度出发，生物体中的微观运动可以分为流动和扩散两种类型。相较于直观的血流在血管中流动，扩散运动不易察觉却无处不在。它可以抽象为微小颗粒在受限环境中的布朗运动（Brownian motion）[40]，能够描述细胞自身的迁移和温度，细胞内的有机分子或代谢产物，以及不同物质在细胞内的运动方式[41]。

当 OCT 探测光与生物组织相互作用时，吸收、单次散射、多次散射等过程同时发生。由于 OCT 成像系统同时利用了相干门和共聚焦门两种机制来抑制多次散射光子，因此可以假设满足一阶玻恩近似（Born approximation）[42,43]，这时 OCT 成像可以近似为系统点扩散函数与微小颗粒散射体之间的卷积[42,43]。根据动态光散射理论[44,45]，由于布朗运动和流动，受限环境中的微小颗粒（散射体）之间的距离随时间持续变化，进而导致散射光的时域强度扰动。当采用相干光源（激光）或部分相干光源（如超荧光源、超连续谱激光等），散射光会在周围颗粒的作用下会经历相长或相消干涉，进而引发更强的扰动。

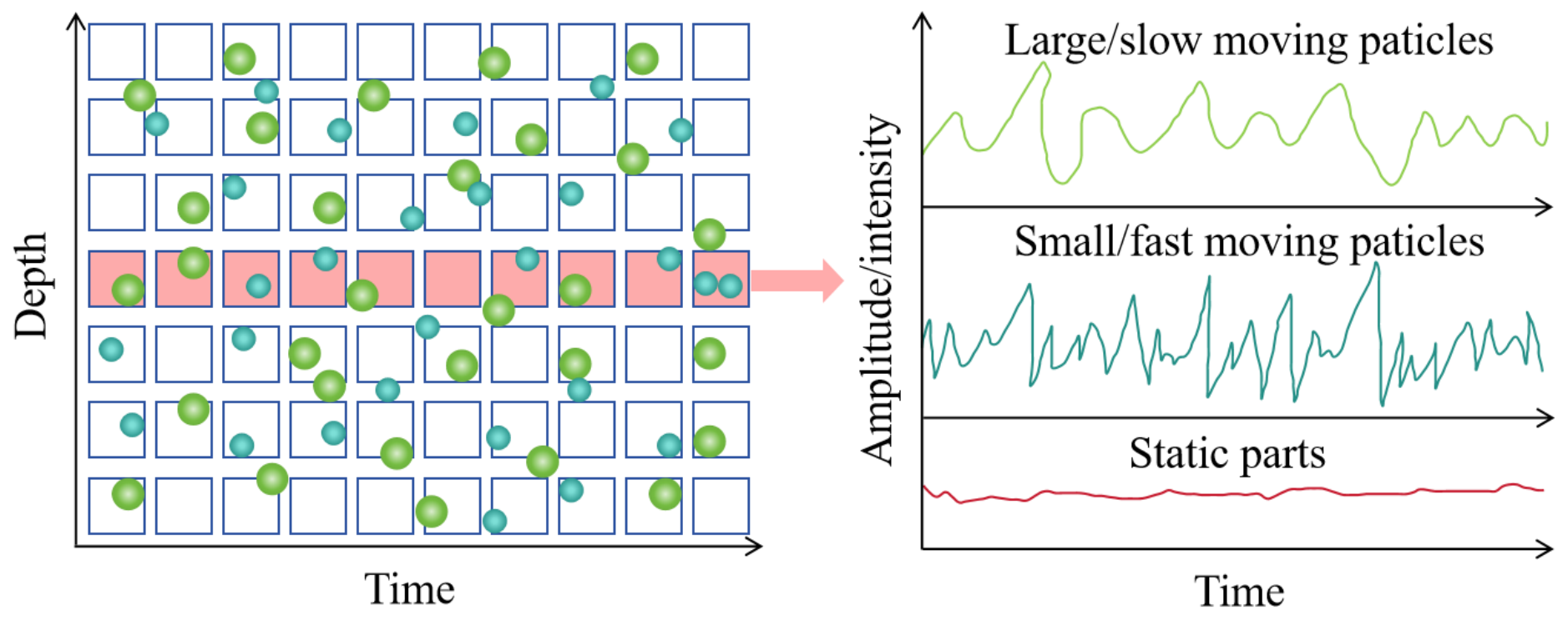

图2 生物体内微观运动带来的OCT信号幅度动态变化示意。

Fig. 2 Illustration of dynamic changes in OCT amplitude due to microscopic movements in an organism.

图 2 给出了生物体内微观运动带来的 OCT 信号幅度动态变化示意。在同一（深度）位置时域连续采样的情况下，大尺寸/缓慢运动的颗粒导致的信号幅度/强度变化大、频率低；小尺寸/快速运动的颗粒导致的信号幅度/强度变化大、频率高；没有颗粒的位置，即静态部

分的信号幅度/强度变化非常小。从以上物理机制出发，可以发展出不同类型的功能成像方法。

上述物理过程也可通过动态散斑（dynamic speckle）现象的相关理论和方法来描述，感兴趣的读者请参考[46–48]。

# 3 基于光场幅度动态变化功能成像的方法

为获取 OCT 光场幅度信号的动态变化，研究者们发展出了不同类型的扫描（时空采样）方式，如图 3 所示。其中主要包括（a）MB 扫描、（b）BM 扫描、（c）双向扫描、（d）逐面体扫描、（e）体间扫描和（f）李萨如（Lissajous）式扫描[49]。

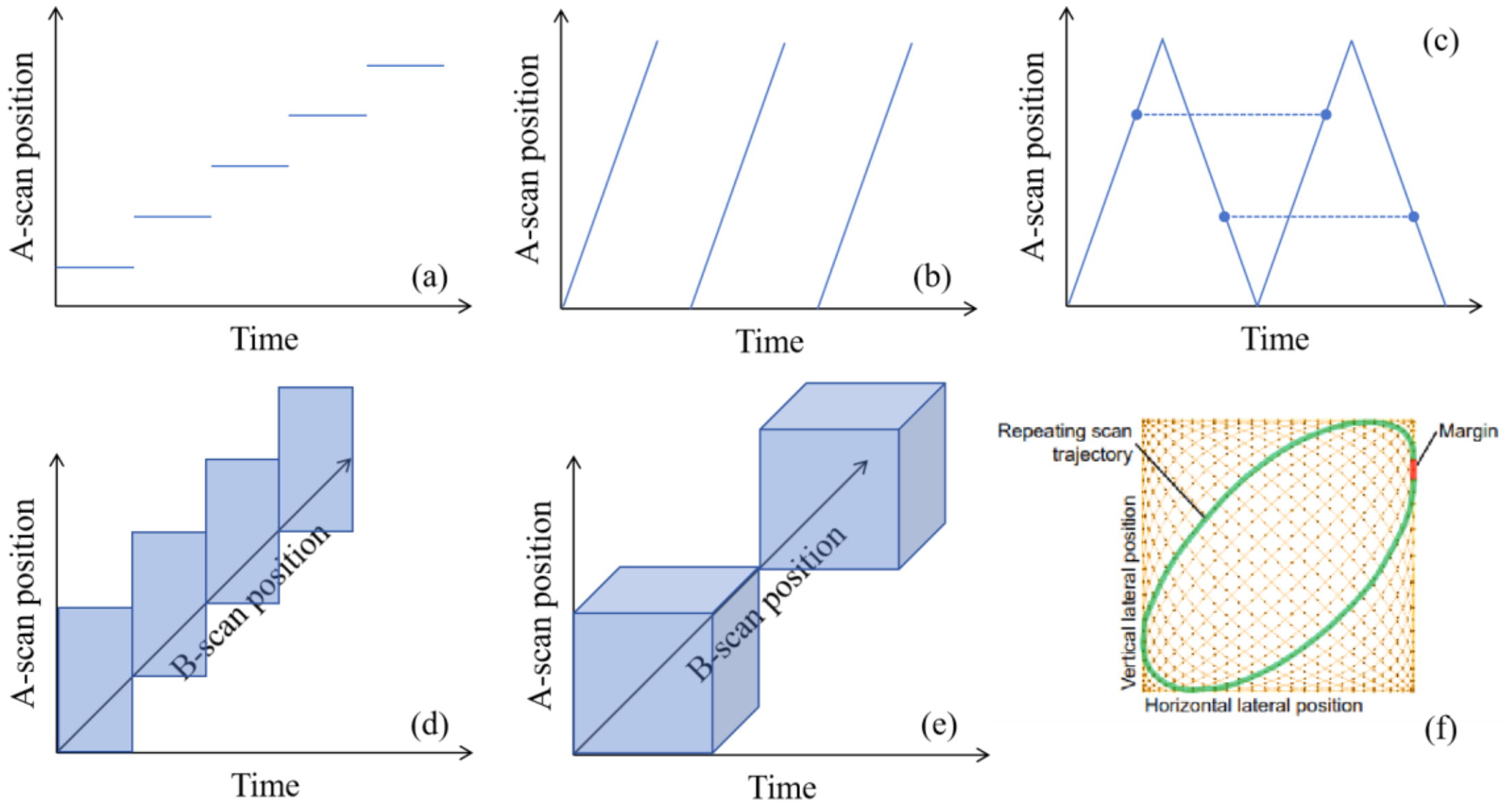

图3 获取OCT光场幅度信号动态变化的不同类型的扫描（时空采样）方式。（a） MB扫描；（b） BM扫描；（c） 双向扫描；（d） 逐面体扫描；（e） 体间扫描；（f） 李萨如（Lissajous）式扫描[49]。

Fig. 3 Different types of scanning (spatiotemporal sampling) protocols for acquiring dynamic changes in OCT amplitude. (a) MB scanning; (b) BM scanning; (c) bidirectional scanning; (d) frame-by-frame volumetric scanning; (e) intervolume scanning; and (f) Lissajous scanning[49].

MB 扫描是指在同一个 A 扫描位置连续获取多个时间信号再移动至下一个 A 扫描位置。BM 扫描则是指连续获取由多个 A 扫描组成的一个 B 扫描，然后再重复获取同一个 B 扫描。为提升 BM 扫描方式的效率，Yifan Jian 等人提出一种双向扫描模式[50]如图 3（c）所示。传统 BM 扫描方式在完成单个 B 扫描后需由扫描元件（通常为振镜）复位到初始位置以进行下一次 B 扫描，这一过程被称为飞回（flyback）。由于飞回过程中获得的数据点并不用于后续的数据处理，为避免浪费，这一过程中通常设置较少的采集点，但给扫描器件的稳定运行

带来了挑战。双向扫描模式通过对称获取扫描元件往返过程中的所有数据点，避免了飞回过程，也确保了扫描器件的稳定性，同时，如图 3（c）中虚线所示的扫描点配对模式可以保证采样时间间隔的一致性。在图 3（a）（b）（c）所示的面扫描重复获取方式的基础之上，可通过连续 B 扫描位置的依次获取，得到三维动态变化，即图 3（d）所示的逐面体扫描方式。另一种三维动态变化的获取方式则是体间扫描，如图 3（e）所示，是直接通过依次获取多个体扫描数据来得到时间维度信息。Yoshiaki Yasuno 等人提出了一种李萨如（Lissajous）式扫描模式[49]，如图 3（f）所示。相较于上述逐行扫描（raster scanning）方式，李萨如扫描更有利于对运动（如眼动）伪影的矫正[51]。此外，还可以采用螺旋线扫描方式以避免扫描元件的急转急停[52,53]。

在 OCT 血流成像和温度层析成像中，为更好的区分组织的静态和动态部分，同一 A 扫描位置的相邻时间采样点间隔通常为毫秒级[54]。现有的 OCT 系统速度通常在十万赫兹左右，因此上述两种功能成像中通常采用 BM 扫描方式。在动态光散射 OCT 和动态 OCT 中，由于需捕获高频信号变化，因此通常采用 MB 扫描方式。

在获取 OCT 光场幅度信号的动态变化后，需要通过合适的数据分析方法以得到相应的功能成像信息。本质上，基于光场幅度动态变化功能成像的数据分析方法可以分为两类，如图 4 所示。一类是图 4（a）所示的基于动态光散射的分析方法，主要被应用于血流成像、三维测速和温度层析成像，另一类是图 4（b）所示的基于频谱的分析方法，主要被应用于组织/细胞特异性显示。

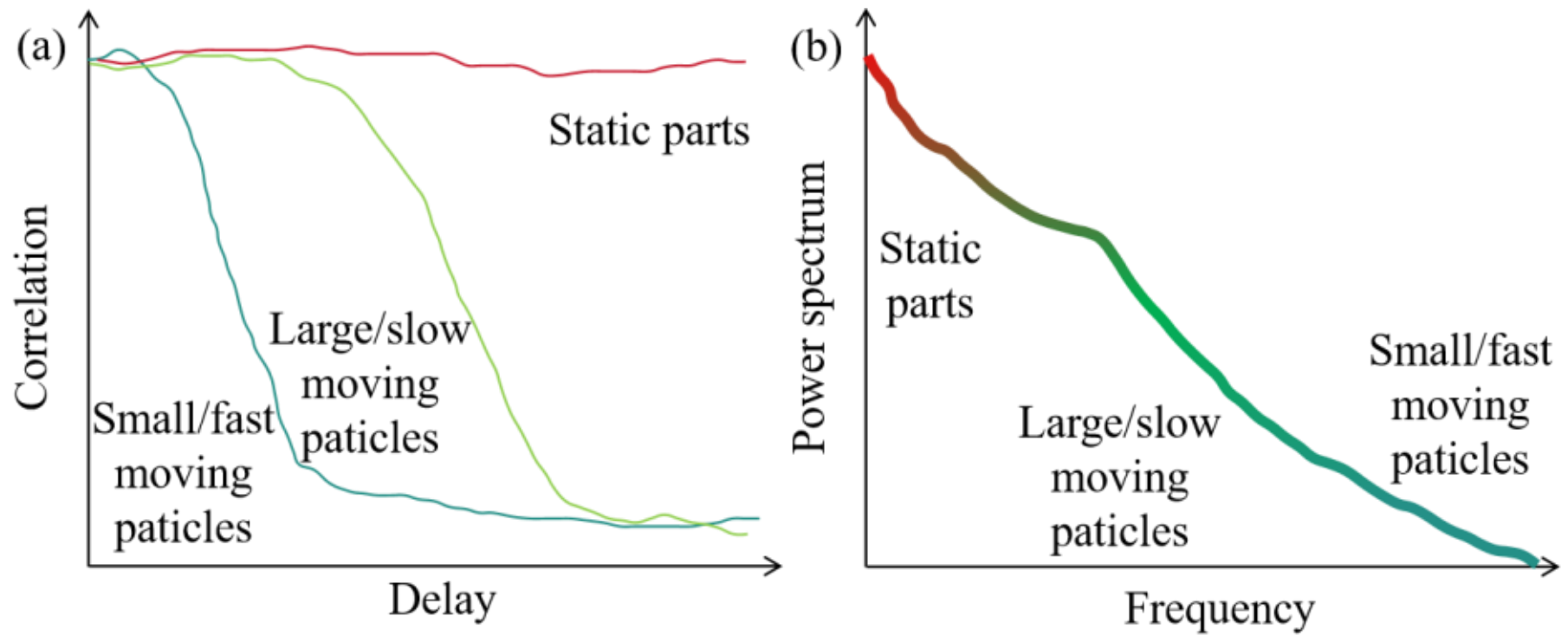

图4 基于光场幅度动态变化功能成像的数据分析方法。（a） 基于动态光散射的分析方法；（b） 基于频谱的分析方法。

Fig. 4 Data analysis methods for extracting functional information from dynamic change of OCT amplitude. (a) dynamic light scattering based methods; (b) power spectrum based methods.

基于动态光散射的分析方法是从信号的时域相关性出发，包括针对散射光场幅度 $E$ 的一阶相干性函数 $g_1$，和针对光场强度 $I$ 的二阶相关性函数 $g_2$（均为归一化形式）：

$$g_1(\tau)=\frac{\langle E(t)E(t+\tau)\rangle}{\langle E(t)\rangle^2},\quad (1)$$

$$g_2(\tau)=\frac{\langle I(t)I(t+\tau)\rangle}{\langle I(t)\rangle^2},\quad (2)$$

其中，$t$ 为时间，$\tau$ 为时间间隔，$\langle\ \rangle$ 表示统计平均。$g_1$ 和 $g_2$ 的关系由西格特关系（Siegert relation）来描述：

$$g_2(\tau)=1+\beta\left|g_1(\tau)\right|^2,\quad (3)$$

其中，$\beta$ 是相干因子，取决于探测器面积、探测光路和颗粒的散射特性。简单考虑单分散颗粒情况下有[55,45]：

$$g_1(\tau)=e^{-2D_\tau q^2\tau},\quad (4)$$

$$g_2(\tau)=1+\beta e^{-2D_\tau q^2\tau},\quad (5)$$

其中，$D_\tau$ 是扩散系数，$q$ 是布拉格波矢[45]：

$$q=\frac{4\pi n}{\lambda}\sin(\frac{\theta}{2}),\quad (6)$$

其中，$n$ 是折射率，$\lambda$ 是入射光波长，$\theta$ 是光场探测器的放置角度。

在动态光散射测量液体微观颗粒属性的应用中[45]，以上公式结合斯托克斯-爱因斯坦方程（Stokes–Einstein equation）可以得到颗粒的半径[56]。在 OCT 应用中 $D_\tau$ 与散射体的运动速度有关[54]。上述内容结合图 2 和图 4（a）可以知道，对于静止部分，其相干性随时间变化不大；对于大尺寸/缓慢运动的颗粒，其相关性随时间间隔增大较慢下降；对于小尺寸/快速运动的颗粒，其相关性随时间间隔增大快速下降。基于光场幅度动态变化的 OCT 血流成像和温度层析成像在时间上只进行了较为稀释的采样，用于区分静态组织和运动部分（血流和热致形变）；三维测速则是进行了稠密的时域采样，基于动态光散射理论发展出速度量化方法。

基于频谱的分析方法则相对较为简单直接，如图 4（b）所示，通过将 OCT 光场幅度/强度变化的时间序列进行傅里叶变换，就可以在频率域上将接近零频的静态部分，低频的大尺寸/缓慢移动颗粒，高频的小尺寸/快速运动颗粒。然后，对不同频率成分进行颜色编码，就可以实现对组织/细胞的特异性显示[57]。此外，基于动态光散射的方法也被应用于组织/细胞的特异性显示，如组织中亚细胞代谢[27],并与基于频谱的分析方法相结合，实现了对视网膜

类器官的三维特异性成像[31]。

# 4 基于光场幅度动态变化功能成像的技术实现和应用

## 4.1 OCT 血流成像

OCT 血流成像在眼底病诊断中取得了巨大的临床转化成功，近年来研究者们已经撰写的大量的相关综述[58,59,32,60–62]，本文仅介绍关键技术方法以及部分最新的研究和应用进展。

OCT 血流成像算法通常可以分为基于相位、基于幅度/强度和复信号的三类[63,34,64,65]。表 1 汇总了主要的基于幅度/强度动态变化的 OCT 血流成像算法，包括 SSADA[21]、svOCT[22] 和 cmOCT[23]。其中、$A$ 和 $I$ 分别为 OCT 探测信号的幅度和强度。$z$ 为深度方向位置，$x$ 为横向位置。$N$ 是重复扫描次数，$M$ 是分光谱的数量。对于平均强度 $\overline{I}$，$\left[A\ B\right]$ 表示其窗函数的尺寸。

表 1 基于光场幅度/强度动态变化的 OCT 血流成像算法

Table 1 OCT angiography algorithms based on amplitude/intensity

| Types of algorithms | Expressions |
| --- | --- |
| SSADA[21] | $1-\frac{1}{M}\sum_{m=1}^{M}\frac{A_{1,m}(x,z)A_{2,m}(x,z)}{\left[0.5A_{1,m}(x,z)^2+0.5A_{2,m}(x,z)^2\right]}$ |
| svOCT[22] | $\frac{1}{N}\sum_{i=1}^{N}\left[I_i(x,z)-\frac{1}{N}\sum_{i=1}^{N}I_i(x,z)\right]^2$ |
| cmOCT[23] | $\sum_{p=0}^{A}\sum_{q=0}^{B}\frac{\left[I_1(x+p,z+q)-\overline{I_1(x,z)}\right]\left[I_2(x+p,z+q)-\overline{I_2(x,z)}\right]}{\sqrt{\left[I_1(x+p,z+q)-\overline{I_1(x,z)}\right]^2+\left[I_2(x+p,z+q)-\overline{I_2(x,z)}\right]^2}}$ |

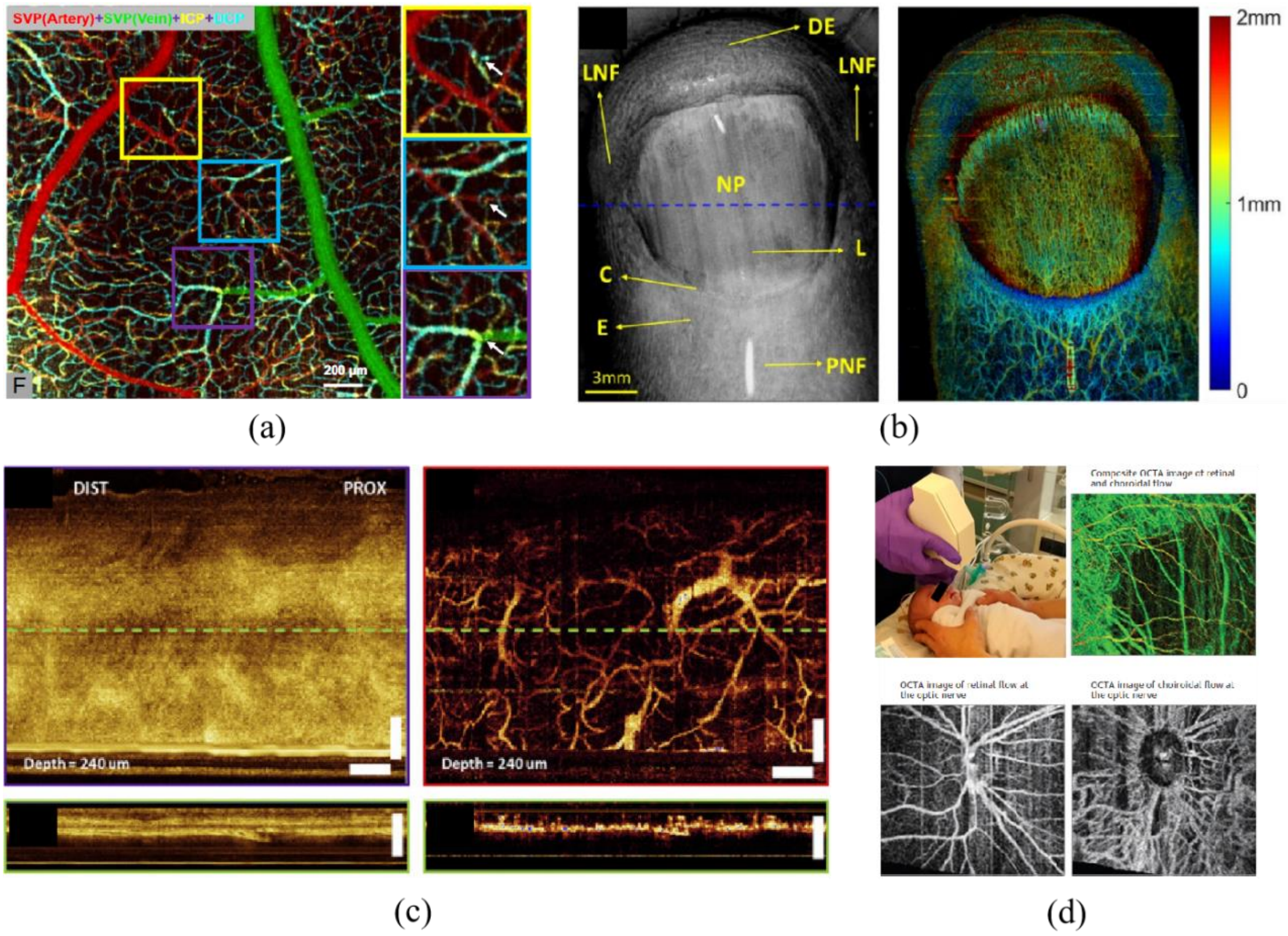

图 5 OCT 血流成像中的新技术和新应用示例。（a） 可见光 OCT 血流成像测量视网膜毛细血管血氧[66]；（b） 超大视场高灵敏度 OCT 用于手指毛细血管成像[67]；（c） 超高速内窥式 OCT 探头获取的消化道皮下微血管结构[68]；（d） 手持式 OCT 血流成像的早产儿视网膜病变诊疗应用[69]。

Fig. 5 Examples of new techniques and applications in OCTA. (a) visible-light OCTA to measure retinal capillary blood oxygenation[66]; (b) ultra-wide-field high-sensitivity OCT for finger capillary imaging[67]; (c) subcutaneous microvasculature of the GI tract acquired by an ultra-high-speed endoscopic OCT probe[68]; (d) Handheld OCTA for the diagnosis and treatment of retinopathy of prematurity[69].

图 5 中给出了一些 OCT 血流成像中的新技术和新应用示例。图 5（a）中，Yali Jia 等人利用可见光 OCT 血流成像技术，对大鼠从视网膜动脉到毛细血管再到视网膜静脉的整个血管转换过程中的血红蛋白含氧量进行体内评估，并展示对吸入氧气浓度变化的生理反应[66]。Ruikang K. Wang 等人发展一种基于新型扫频激光器的超宽视场 OCT 血流成像系统。该扫频激光器具有非常窄的瞬时线宽和很高的相位稳定度，能够在长达 46 毫米的范围内进行成像。图 5（b）显示该成像系统能够获得具有丰富细节的人体手指血流灌注。James G. Fujimoto 等人使用一个轴向扫描速率为 600 kHz 的内窥镜 OCT 系统[70]，实现了对人消化道的皮下微血管网络血流得成像，这将有助于提高对胃肠道恶性肿瘤前疾病的检测能力，还有利于研究其他血管疾病，如放射性直肠炎、胃窦血管异位、缺血性结肠炎以及肿瘤相关的血管生成[68]。本文作者与合作者们发展出了手持式 OCT 血流成像系统[71]，以解决婴幼儿无法采用标准的

荧光造影进行血流灌注评估的临床痛点。系统被成功应用于全球儿童失明的主要病因——早产儿视网膜病变（ROP）的临床研究中，由图 5（d）可见，它能够有效评估经过激光治疗后的 ROP 血流灌注分布[69]。

### 4.2　动态光散射 OCT 和其三维测速应用

动态光散射 OCT 是动态光散射理论[45,44]和 OCT 探测方式的结合，由 David A. Boas 等人首先提出[26]，旨在实现对异质性扩散和流动的高分辨率三维成像。相较于多普勒 OCT 仅对血流的轴向分量敏感，动态光散射 OCT 能够同时实现对血流的三维方向分量的测量[72,30]。相较于 OCT 血流成像只能区分血流和静态组织[35,73]或进行半定量的流速测量[54,74]，动态光散射 OCT 则能够精确测定血流的流速和扩散系数[26,75,76]。

早期的动态光散射 OCT 的信号分析中采用了复光场相关函数[26] $g(\tau)$ [26]：

$$g(\tau) = M_S + M_F e^{-h_t^2 v_t^2 \tau^2 - h_z^2 v_z^2 \tau^2} e^{-q^2 D_\tau \tau} e^{iqv_z \tau} + (1 - M_S - M_F)\delta(\tau), \qquad (7)$$

其中，$M_S$ 和 $M_F$ 分别是静态部分和血流在组织中所占的比例。$h_z$ 和 $h_t$ 分别是轴向和横向空间分辨率的倒数。$v_t^2 = v_x^2 + v_y^2$ 是血流速度的横向面分量，$v_z$ 是血流速度的轴向分量。$\delta(\bullet)$ 是狄拉克（Dirac）$\delta$ 函数（其它参数的定义见第 3 章）。测量得到的相关函数通过拟合算法可以得到 $M_S$、$M_F$、$v_t$、$v_z$ 和 $D$ 五个参数[25]。

最近，Konstantine Cheishvili 等人考虑到相位测量对成像系统的更高要求，发展出仅使用光场幅度变化信息的、基于二阶相关函数 $g_2(\tau)$ 的动态光散射 OCT 方法。同时提出采用 B 扫描以改进 M 扫描中的测量速度限制[75]。其相关函数可以写作：

$$g_2(\tau, z) = A_2(z) e^{-2D_\tau q^2 \tau} e^{-\frac{(v_0(z)-v_s)^2 \sin^2\theta\tau^2}{w_z^2}} e^{-\frac{2(v_0(z)-v_s)^2 \cos^2\theta\tau^2}{w_0^2}}, \qquad (8)$$

其中，$A_2(z)$ 是深度方向上与信号强度相关的因子。$v_0(z)$ 是总流速，$v_s$ 是沿着血流方向的有效扫描速度。$\theta$ 是血流方向与横向面的夹角。$w_0$ 是 OCT 探测高斯光束的束腰半径，$w_z$ 是相干函数的束腰半径。相较于多普勒 OCT 和采用 M 扫描模式，该项研究中的结果表明新方法可以在各种血流方向下获得更大的速度测量范围和更高的测量精度，如图 6（a）所示。

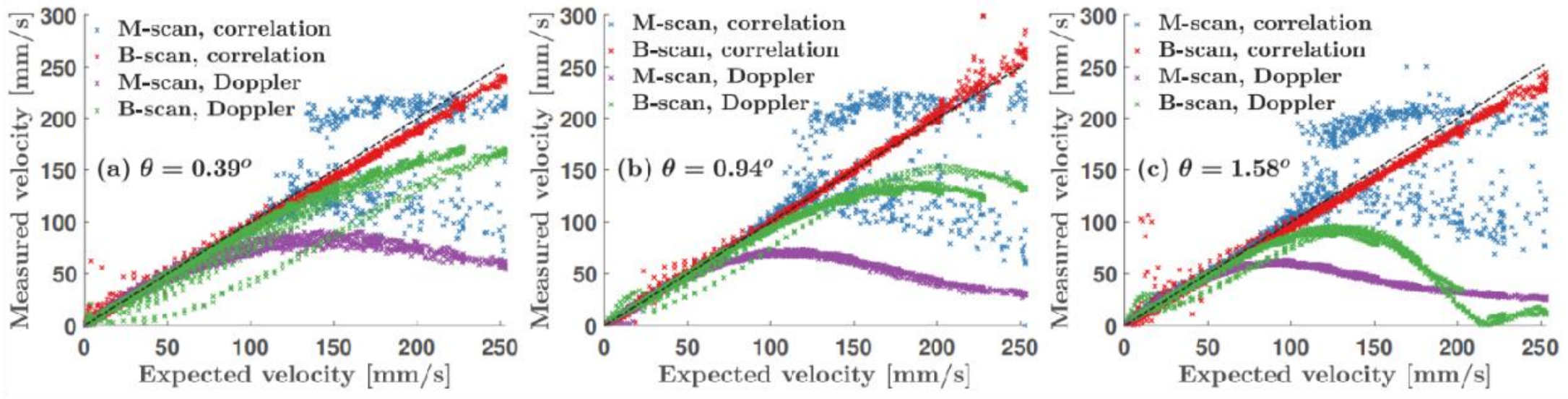

(a)

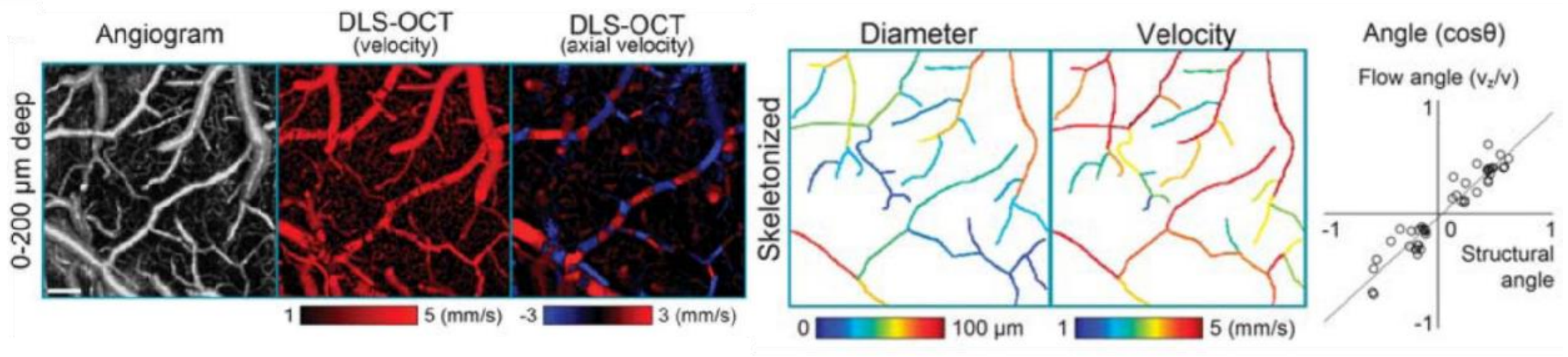

(b)

图 6 动态光散射 OCT 的三维血流测速应用示例。 （a） 采用 B 扫描模式的动态光散射 OCT 测量血流速度（红色）与多普勒 OCT 和 M 扫描模式下的结果对比[75]; （b） 脑血管血流速度定量成像应用[77]。

Fig. 6 Measurement of blood flow velocity using dynamic light scattering OCT in B-scan mode (red) compared with results in Doppler OCT and M-scan mode[75]. (b) quantitative imaging of cerebral blood flow velocity[77].

图 6（b）是动态光散射 OCT 在脑血管血流速度定量成像应用示例[77]。该项研究通过对大脑皮层获得的相关函数数据进行分析，得到了绝对速度、轴向速度和扩散系数的三维图。其中，绝对速度图中观察到的血流模式与传统 OCT 血管造影中的血流模式非常接近，动态范围较大，可以定量测量动脉和静脉中的 CBF 速度。轴向速度图则与传统的多普勒 OCT 图像相似。血流角度与血管造影观察到的血管结构角度一致。

此外，还有一类方法被称为 OCT 毛细血管血流测速（capillary velocimetry），其方法原理与动态光散射 OCT 类似，但具体的技术实现则有不同。本文不对其做进一步叙述，感兴趣的读者请参考[78–80]。

## 4.3 动态 OCT 和其组织/细胞特异性成像应用

动态（dynamic）OCT 这一名称，除了被用于组织/细胞特异性成像外，还被用于皮肤的 OCT 血流成像、角膜形变测量等研究领域[81,33,82,83]，但它们不在本文的讨论范围内。此外，本文只是从基于光场幅度动态变化的宏观角度对这一技术进行简述，更具体技术和应用细节请参考近期的综述文章[84]。

动态 OCT 的兴起首先受益于近年来时域全场（full-field）OCT 的发展[85,86]，其相较于

传统显微镜具有显著的深度方向分辨能力优势。随后，Michael Münter 等人用点扫描的频域 OCT 也实现了动态 OCT 成像[57]，证明了该技术具有良好的 OCT 硬件兼容性。动态 OCT 通过获取组织和细胞位置的幅度动态变化信息以实现无标记的特异性成像，其技术方法已经在第 3 章中给出了概括性的描述，在这里通过示例做进一步的说明。

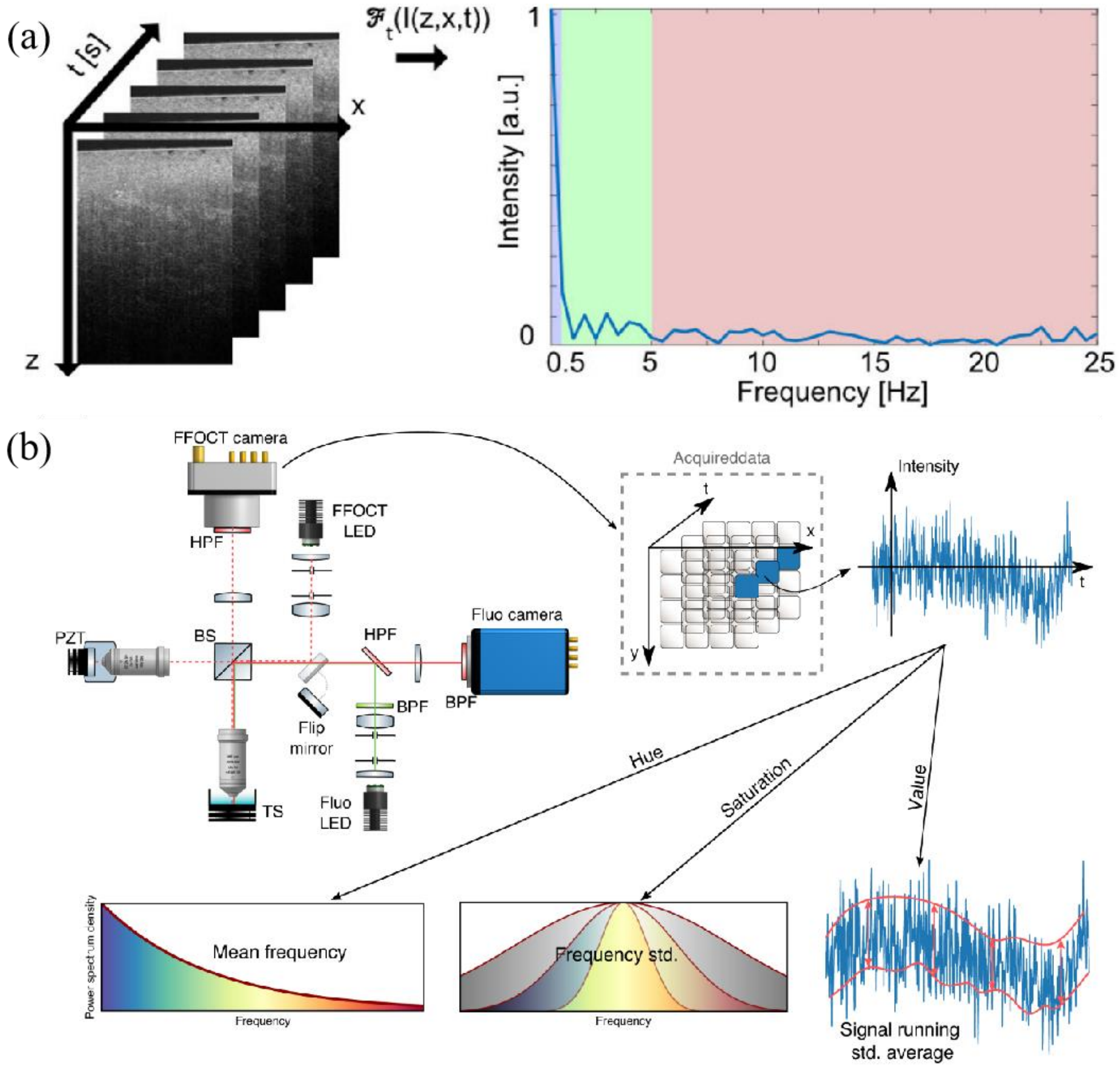

图 7 动态 OCT 的技术实现方案示例。（a） BM 扫描模式+基于频谱的分析方法[57]；（b） 体间扫描模式+结合了基于频谱和基于动态光散射的分析方法[31]。

Fig. 7 Examples of technical realization schemes for dynamic OCT. (a) BM scanning mode + spectrum-based analysis method[57]; (b) intervolume scanning mode + combination of spectrum-based and dynamic light scattering-based analysis method[31].

图 7 中给出了动态 OCT 的两种具体技术实现方案。图 7（a）是利用 BM 扫描模式进行采集，然后使用了基于频谱的分析方法[57]。该项研究对每个体素评估 OCT 信号绝对值的时间变化。对时间序列进行傅立叶变换，计算三个频段的积分振幅。像素在 RGB 图像中以颜色编码，代表运动活动的不同时间尺度。蓝色代表慢速运动（0-0.5 Hz），绿色代表中速运动（0.5-5 Hz），红色代表快速运动（5-25 Hz）。图 7（b）是利用体间扫描模式进行采集，

然后结合了基于频谱和基于动态光散射的分析方法[31]。该项研究采用了 HSV 而非 RGB 色彩空间，使得每个通道具有便于视觉解释的物理属性。其中，H 通道是整个功率谱的平均频率，S 通道是频率带宽的倒数，V 通道则是随时间变化信号的标准差，其定义为[27,84]：

$$V(x,y)=\left\langle \sqrt{\frac{1}{N}\sum_{i=1}^{N}\left(I(x,y,t_i)-\langle I(x,y)\rangle\right)^2} \right\rangle, \quad (9)$$

其中，$(x,y)$ 是 OCT 体扫描 en face 平面上的位置，$t=1\ldots N$ 是时间维，$I$ 是 OCT 信号强度。

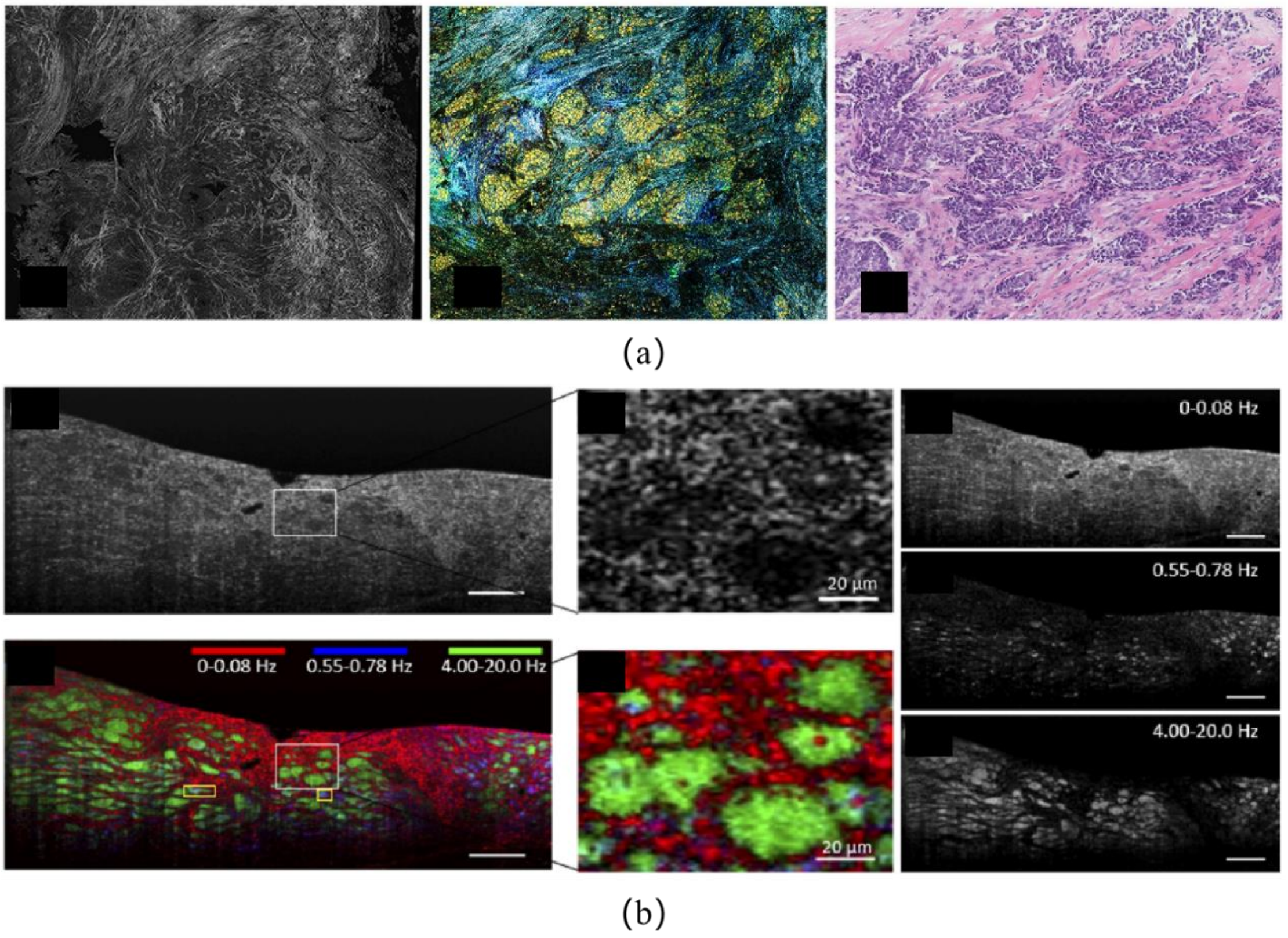

图 8 动态 OCT 的组织/细胞特异性成像应用示例。（a） 浸润性乳腺癌组织[87]；（b） 食管组织[88]。

Fig. 8 Examples of tissue/cell-specific imaging applications of dynamic OCT. (a) Infiltrating breast cancer tissue[87]; (b) esophageal tissue[88].

图 8 是动态 OCT 的组织/细胞特异性成像应用示例。Shu Wang 等人将 OCT 强度和动态 OCT 应用于正常和癌变乳腺组织、良性乳腺病变以及切除的腋窝淋巴结。然后对组织进行常规处理和染色（苏木精-伊红），以进行比较[87]。图 8（a）是该项研究中对浸润性乳腺癌组织的成像结果。可以看到癌细胞的侵袭在 OCT 强度图中表现为局灶性或圆形低密度，在动态 OCT 中则表现为易于区分的恶性细胞（即大而亮的黄色细胞，核染色深）。Guillermo

J. Tearney 等人在之前研发的 1 微米分辨率 OCT 系统[89]的基础上应用了动态 OCT 的分析方法，对新鲜切除的人体食管和宫颈活检样本进行了成像[88]。图 8（b）是该项研究中对一位因慢性胃食管反流病而接受内窥镜检查的患者的食管活检图像。OCT 结构图像中的折射率对比度不足，且散斑的存在进一步阻碍了细胞特征的详细观察。使用动态 OCT 后细胞对比度大大提高。由于细胞质中的光散射运动比细胞膜和细胞间隙的运动频率高，细胞核等细胞器的频率较低，使用动态 OCT 能够有效地将它们区分开来。

### 4.4 OCT 温度层析成像和其热物理治监应用

OCT 温度层析成像的主要实现方式是通过 OCT 血流成像中的基于动态光散射的分析方法，如散斑方差、散斑去相关等[22,90]，来提取热致组织形变中的微观运动，进而建立其与温度之间的关系[91,28,92–94]（除基于光场幅度动态变化的技术方案外，还有少数基于偏振态、相位等的 OCT 温度层析成像研究[95–97]，但它们不在本文的讨论范围内）。

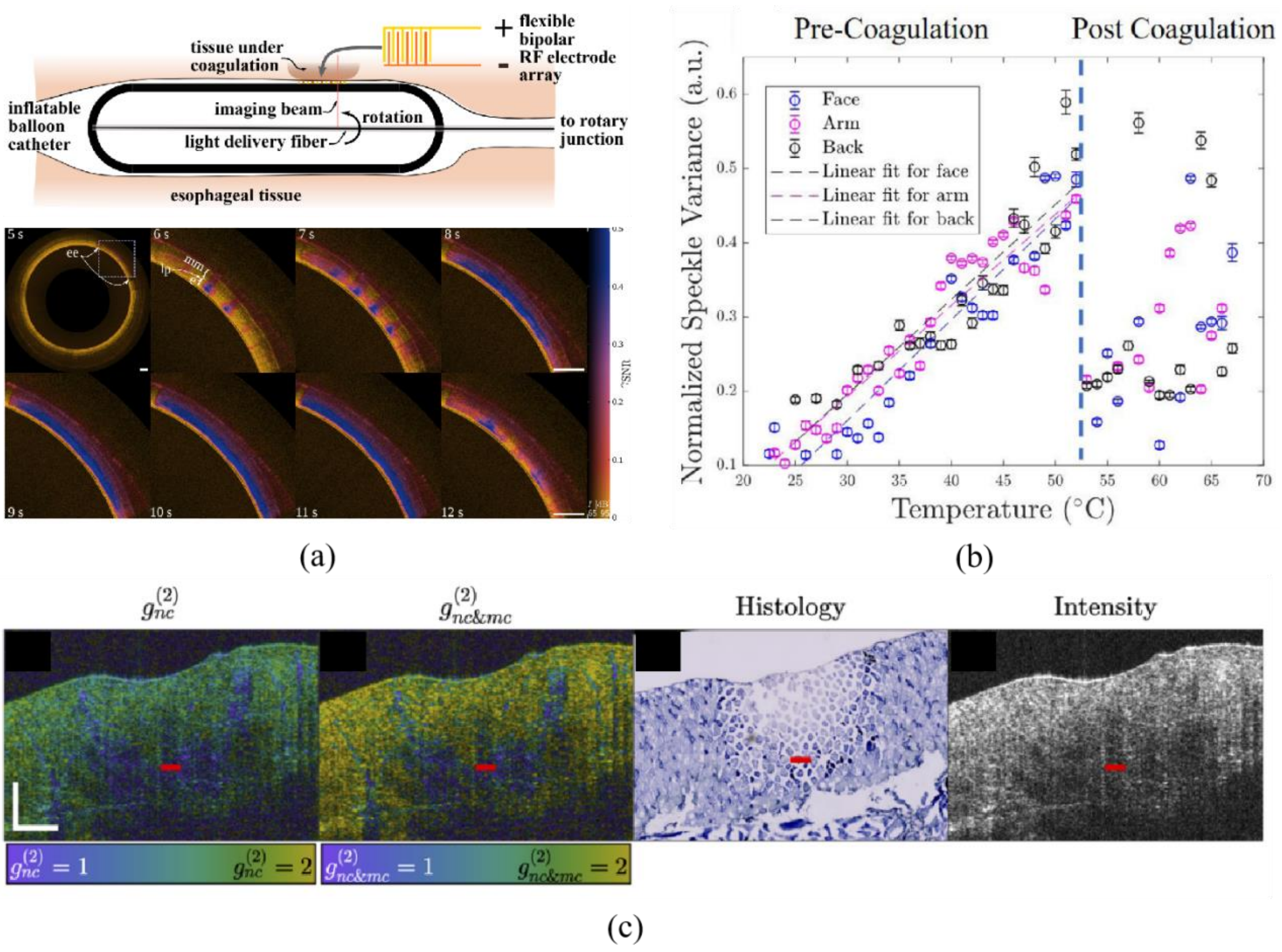

图 9 OCT 温度层析成像应用于热物理治疗监控的技术和效果示例。（a） 食道球囊导管射频消融温度场监控[92]；（b） 激光治疗皮肤病灶时散斑方差信号与温度之间的关系[28]；（c） 应用了噪声矫正散斑强度去相关算法的激光治疗深度显示[91]。

Fig. 9 Techniques and results of OCT thermometry applied to thermophysical therapy monitoring. (a) RF ablation temperature field monitoring of an esophageal balloon catheter[92]; (b) relationship between the speckle variance signal and temperature during laser treatment of skin lesion[28]; and (c) visualization of the depth of laser treatment using a noise-corrected speckle decorrelation algorithm[91].

图 9 给出了一些 OCT 温度层析成像应用于热物理治疗监控的技术和效果的例子。Brett E. Bouma 等人应用 OCT 监控射频球囊导管治疗巴雷特食道病灶[92]，利用复微分方差算法[98]能够有效显示出治疗热场的范围，且与病理切片的结果相一致。图 9（a）上部分显示射频治疗球囊的结构和 OCT 成像耦合方式，下部分是治疗过程中的 OCT 图像时间序列。可以从中清晰的看到治疗热场的变化情况。Jin U. Kang 等人研究了使用散斑方差 OCT[22]来监测激光照射过程中被切除的人体皮肤组织样本的实时温度变化[28]。结果显示该方法能够测量激光治疗过程中组织温度的快速变化。同时，研究发现在组织凝固前散斑方差信号与温度间具有线性关系，在凝固后则线性关系失效，如图 9（b）所示。Raphaël Maltais-Tariant 等人介绍了一种能够通过双包层光纤的 OCT 成像和监控激光治疗的系统[91]。其中使用了一种新型的矫正导致去相关的解析算法[90]，能够实时识别热凝固的深度，并在达到目标深度时自动关闭治疗激光。如图 9（c）所示，相较于 OCT 强度信号，采用了噪声矫正的散斑去相关的算法能够准确显示治疗范围并识别治疗深度（图中的红色短线所示）。

# 5 基于光场幅度动态变化功能成像的技术挑战和未来展望

相较于相位、偏振态、波长等其它光场物理量，幅度的获取对 OCT 成像系统的硬件和算法的要求最低。如上所述，通过在幅度的测量中加入时间维度，能够获取生物组织结构之外的大量功能信息，如血流、血氧、流速、细胞特异性、温度等。其中，OCT 血流成像已经在临床上加入诸多眼底病的标准诊断流程，如青光眼、糖尿病视网膜病变、老年性黄斑病变等。动态光散射 OCT、动态 OCT、OCT 温度层析成像等已经展现出了良好的应用前景，有潜力进一步推动 OCT 的定量化和多功能化，在智能、精准医疗中发挥作用。

与此同时，时间维度的引入也意味着成像过程的加长，这为上述功能成像的在体应用带来了挑战。第 2 章中阐明了基于光场幅度动态变化功能成像本质上是在测量生物体内的微观运动。然而，生物体，特别是人体，还存在着诸多更大尺度的运动，包括身体不自主的晃动、眼球微动、肌肉组织施力时的颤动、心跳等[99–101]。这些运动都是 OCT 光场幅度动态变化测量中的噪声来源。即使是已经被临床采用、在时间维度上采样非常稀疏的 OCT 血流成像，仍需要体噪声消除、运动伪影矫正、配准等多种类型后处理算法[102–105]，扫描激光检眼

镜、双目相机等眼球追踪部件[106–108]，和颌托等辅助稳定装置[109]。对于需要稠密时域采样的动态光散射 OCT、动态 OCT 等，当它们发展到在体研究阶段时，这一问题将会更加严重，需要更多的解决运动噪声的思路。例如，Gijs van Soest 等人发展出 heartbeat OCT 技术[110]，通过心电图触发 OCT 采集系统使其与心动周期同步，从而消除了血管内 OCT 成像中由心跳引起的运动噪声。

运动噪声的减少也可通过提高 OCT 系统的采集速度，或降低数据分析中对时间采样点数的需求。Robert A. Huber 等人通过发展傅里叶锁模激光器将 OCT 成像系统的 A 扫描速度提升至数个兆赫兹[111]。Ping Xue 等人通过对皮秒超连续谱激光器的时域拓展和缓冲进一步将 A 扫描速度提升至 40 兆赫兹[112,113]。在商业化系统中，基于短腔和垂直腔面发射激光器（VCSEL）以及微电机系统（MEMS）滤波的 100 至 400 千赫兹扫频光源 OCT 已经成为了主流[114,111]。谱域 OCT 系统受益于高速 CMOS 线阵相机的发展，也已达到了 250 千赫兹的 A 扫描速度[115]。另一方面，通过主成分分析（PCA）等数据降维算法可以在减少数据点数量的情况下量化流速，Benjamin Vakoc 等人已经在基于 OCT 强度信号的血流速度分析中进行了证明[116]。随着近年来人工智能特别是深度学习技术的高速发展[117]，采用更为强大的数据降维算法进行 OCT 时域信号分析有着巨大的发展空间。此外，OCT 血流成像通过时间维度的稀疏采样能够获得鲁棒的无标记血管造影，虽然牺牲了对绝对流速的测量，但仍可定性/半定量的表征血液动力学[118–120]。例如，Peng Li 等人通过分析 OCT 血流成像信号，提取到神经刺激之后的血流动力学响应，并在糖尿病模型中观察到了神经血管响应功能的异常[119,120]。

在进行深度信息层析的同时，横向面上的信息获取对于 OCT 功能成像的生物医学应用同样至关重要，即所谓的 en face OCT[86]。其中的核心预处理步骤——B 扫描图像的自动分割问题，近年来已由以 U-Net 为代表的深度学习算法较好的解决[121–123]。然而，en face 面上的采样点数量/密度问题仍然阻碍着基于光场幅度动态变化功能成像。如上所述，由于时间维度采样的需求，成像过程已经相较于常规的 OCT 结构成像大为增长，en face 面的采样将进一步加重图像噪声、伪影等问题。但信息的完整获取需满足奈奎斯特-香农采样定律（Nyquist-Shannon sampling theorem）[124]，意味着在 OCT 中，横向采样点间隔需是横向光学分辨尺寸的一半（通常在 10 微米以下）。与此同时，许多疾病（如糖尿病视网膜病变）的诊断需要获取大视场以避免漏检。这一矛盾导致了上述功能成像中信息的失真。如在 OCT 血流成像中，已有多项研究发现大视场成像下的生物标记物（如血管密度、血流指数等）失准[125–127]，这是由于商业化 OCT 系统在进行大视场扫描时未能保持所需的采样密度。这一

问题同样可以随着 OCT 成像系统的速度提高而得到缓解，近年来生成式深度学习和压缩感知技术在其中的应用也初见曙光[128–130]。

在点扫描 OCT 之外，线场（line field）OCT 和全场（full field）OCT 也被用 OCT 结构和功能成像中[131,85]。对于谱域和扫频光源线场或全场 OCT 而言，虽然（部分）牺牲了共聚焦条件，但大幅提升了成像速度，对于生物医学中的动态过程记录和光场变化的高频信息记录大有裨益[132,133]。然而，在当前的硬件技术水平下，此类系统的搭建成本显著高于通常的点扫描 OCT 系统，因此仍被较少采用。另一方面，时域全场 OCT 由于其较低的硬件成本和 en face 平面高速、高分辨率成像在动态 OCT 中被大量采用。其中通常采用林尼克（Linnik）干涉仪构型，在在参考臂和样品臂上同时使用显微镜物镜，并配备了高速压电平移台用于光程控制[84,134]，与之配备的二维相机通常需要 80 Hz 以上的帧率[84]。与点扫描动态 OCT 中通常采用的 MB 和 BM 扫描方式不同，时域全场 OCT 是依次记录不同深度的 en face 平面的光场幅度动态变化[84]。

此外，上述基于光场幅度动态变化功能成像技术的理论基础仍待进一步夯实，特别是动态 OCT 和 OCT 温度层析成像仍处于非常早期的发展阶段，方法学中存在大量的经验性操作，这将会阻碍技术的定量化发展和应用可拓展性。动态光散射 OCT 技术由于处理步骤相对较为繁复，为其在研究社区中的推广带来了门槛，期待出现类似于结构光显微镜、光声成像研究社区中开源工具[135,136]。

尽管仍存在上述技术挑战，基于光场幅度动态变化的 OCT 功能成像技术已经在血流成像临床应用中取得了巨大的成功，并在三维流速测量、组织/细胞特异性显示、温度层析成像等方面展示出突出的应用潜力。在眼科、心血管、皮肤科 OCT 影像系统已经大量商业化的今天，基于光场幅度动态变化的 OCT 功能成像仅通过算法和采集模式创新，就能显著拓展该技术的临床应用场景。例如，通过与硬质和柔性 OCT 内镜结合，可在呼吸道、消化道、宫颈等部位实现原位、快速“活检”；通过与穿刺针、球囊等治疗器械以及微波、射频等治疗手段结合，将有助于实现肿瘤、斑块等的诊疗一体化；通过与手术显微镜和机器人结合，将推动治疗过程的精准化和自动化。综上，本文通过对基于光场幅度动态变化的 OCT 功能成像原理和应用进行简要综述，希望能引起相关领域研究者的兴趣，进而共同推动其技术发展和临床转化。

# 参考文献